\documentclass[aps,prl,twocolumn,groupedaddress,showpacs,showkeys,floatfix]{revtex4}

\usepackage{amssymb,amsmath,amstext,graphicx,epsfig}

\begin{document}


\title{Large-scale Reverse Engineering by the Lasso}


\author{Mika Gustafsson}
\email{mikgu@itn.liu.se}
\author{Michael H\"ornquist}
\email{micho@itn.liu.se}
\author{Anna Lombardi}
\email{annlo@itn.liu.se}
\affiliation{Department of Science and Technology, 
	Link\"oping university (Campus Norrk\"oping), 
	SE-601 74 Norrk\"oping, Sweden}


\date{\today}

\begin{abstract}
We perform a reverse engineering from the ``extended Spellman data'', 
	consisting
	of 6178 mRNA levels measured by microarrays at 73 instances
	in four time series during the cell cycle of the yeast 
	\emph{Saccharomyces  cerevisae}.
	By assuming a linear model of the genetic regulatory network, 
	and imposing an extra constraint (the Lasso),
	we obtain a unique inference of coupling parameters.
	These parameters are transfered into an adjacent matrix, which is
	analyzed with respect to topological properties and 
	biological relevance.
	We find a very broad distribution of outdegrees in the network, 
	compatible with earlier
	findings for biological systems and totally incompatible
	with a random graph, 
	and also indications of modules in the network.
	Finally, we show there is an excess of genes coding for 
	transcription factors
	among the genes of highest outdegrees,
	a fact which indicates that our approach has biological relevance.
\end{abstract}


\maketitle

Advances in microarray technologies make it today possible to
measure mRNA-levels for thousands of genes simultaneously.
Also, large-scale measurements of protein levels are gradually 
becoming feasible, as well as results on two-hybrid measurements 
on protein-protein interactions and genome-wide data for DNA binding proteins.
These processes have emphasized the need for computational biology
in order to get as much information out of such measurements
as possible.

One way to handle these data is to infer, or reverse engineer,  genetic
regulatory networks from temporal data.
Although still somewhat speculative, researchers are exploring the
boundaries for what kind of inference that is possible.
There are many approaches for network formation, ranging from Boolean
circuits to very complicated non-linear spatial models,
see \cite{Jong-02} and references therein.
Most models use only transcript data, whereas some incorporate
other chemical constituents as well.
A model  based on mRNA-data only is nothing
but an effective network of gene-to-gene interactions.
This might
look too simplistic in view of the complete network, which includes
metabolites, proteins, etc., but it can be thought of as a projection
onto the space of genes only.
By focusing only on transcript data, 
the networks obtained are not biochemical regulatory
networks, but phenomenological networks where many physical connections
between macromolecules might be hidden by short-cuts, i.e., 
many intermediate units in regulatory cascades might be hidden
\cite{Brazhnik-02}.

A special class, which has gained some popularity, is the
linear, continuous model.
Of course, no one claims there is a linear relationship between
the units in a real regulatory network.
Instead, the working hypothesis is that linear equations can
at least capture the main features of the network.
The main argument is that many functions
can be quite accurately approximated around a specific working point with
their linearization. 
Thus, it can provide a good 
starting point for further considerations.

A key problem for all models are, however, shortage of data.
The number of genes is in general much larger than the number
of measurements, and different authors have taken somewhat
different avenues to remedy this obstacle.
For the linear continuous model based on transcript data,
the first study we are aware of was by D'haeseler \emph{et al.}
\cite{Dhaeseleer-99} and focused on a subset of less than hundred 
genes that were believed to be interrelated.
Their problem was still underdetermined, and they
interpolated the data in order to achieve more, simulated,
measurements.
However, more measurements in the same time-series is an ineffective way of
increasing the information content in the data \cite{Hornquist-03a}.
Another early study was by vanSomeren \emph{et al.} \cite{vanSomeren-00},
who clustered genes
into the same number
of groups as they had measurements, and thus obtained a 
mathematically well-posed problem.
Still another approach was explored by Holter \emph{et al.} \cite{Holter-01},
who formed networks among the principal components of the data.
A more biologically motivated study was performed by Yeung \emph{et al.}
\cite{Yeung-02}, who assumed that the resulting network should 
be sparse and that way got a unique solution.
Somewhat in the same spirit, vanSomeren \emph{et al.} \cite{vanSomeren-03}
conducted a systematic study on how to incorporate prior knowledge
into the inference procedure.
Finally we mention the only large-scale inference we are aware of.
It was conducted by Dewey and Galas 
\cite{Dewey-01}, who considered the whole genome of yeast with more than
6000 genes, and formed the network by taking the solution which
minimized the $L^2$-norm of the coefficients 
and set to zero all matrix elements below a certain threshold. 
However, the resulting network had connections only for 143 genes,
and although they justify that their result makes biological sense 
on a small scale, they lack a large-scale analysis.

In the present letter, we utilize one statistical method, the Lasso
\cite{Tibshirani-96}, to reverse engineer a network among ORFs 
(``Open Reading Frames´´, hereafter referred to as ``genes'') 
in the so-called extended Spellman dataset. 
This dataset  is one of the most referenced sources of
microarray data and 
 contains measured mRNA levels of 6178 genes for the yeast
\emph{S. cerevisae}, presented as logarithms of the fraction
between the measured level and a reference level.
The measurements of interest for us are carried out through
one or more periods of the cell cycle in four time 
series---Alpha, CDC15, and Elutrition from \cite{Spellman-98}, 
and CDC28 from \cite{Cho-98}---with  different
synchronization procedures.
The total number of experiments in all series is 73, divided as
18, 24, 14 and 17 microarray experiments for each series.

The missing data in this set are estimated by the procedure proposed
in \cite{Troyanskaya-01}.
Essentially, it consists of selecting genes with expression profiles 
similar---in the Euclidean distance---to
the gene of interest to impute missing values.
The number of neighboring genes
used to estimate the missing values is here 15.
Finally, we center and normalize the
expression data to have zero mean and unit variance.

We consider
a linear, continuous model of the form
\begin{equation}
\frac{dx_i}{dt}(t)=\sum_{j=1}^N w_{ij}x_j(t) + \epsilon_i.
\label{eqnsofmotion}
\end{equation}
Here $x_i(t)$ denotes the logarithm of the ratio values of mRNA of gene $i$
at time $t$, and $N=6178$ denotes the number of genes.
The coefficient $w_{ij}$ is the effect of gene $j$ on
gene $i$ and does not depend on time.

The network is inferred by minimizing the residual sum of squares,
with an extra constraint on the $L^1$-norm of the coefficients
(the Lasso).
The hyperoctahedronal form of the constraint makes it more
likely that coefficients should become identical zero.
Explicitly, it takes the form
\begin{eqnarray}
\hat{w}_{ij}=
\underset{w_{ij}}{\text{arg min}} & & \sum_{k=1}^K \left(
		\sum_{j=1}^N w_{ij}x_j(t_k) - \frac{d x_i}{dt}(t_k) \right)^2
	\label{lassoobjective}\\
	\text{subject to} & & \sum_{j=1}^N |\hat{w}_{ij}| \le \mu_i \quad
		\text{for} \,i=1,\cdots, N.
\label{lassoconstraint}
\end{eqnarray}
Each microarray measurement is here supposed to have been performed
at time $t_k$ and   there are  $K$ experiments.
The time derivatives are obtained by spline interpolation of the
original data, where we make use of so-called taut splines \cite{deBoor}
in order
to achieve curves that are faithful to the measured data but 
still do not oscillate too wildly.
By this procedure, the problem factorize and we can perform
the minimization for each gene $i$ separately.

For small enough 
constraint parameters $\mu_i$, the solution is unique.
To choose these parameters, we first solve the combined
minimization problem of the residual sum of squares in 
(\ref{lassoobjective}) together with the minimization of the
$L^2$-norm of the coefficients.
These values,
\begin{equation}
\mu^{(2)}_i = \left(\sum_{j=1}^N \left(\hat{w}_{ij}\right)^2\right)^{1/2}
\quad \text{for $i=1,\cdots, N$},
\end{equation}
are used as base-lines against which
we measure the size of the $L^1$-constraints.
Here we  utilize the values $\mu_i=0.1 \mu^{(2)}_i$.
However, varying the coefficient from the value 0.1
does not result in  large changes in the chosen subsets. 
With this choice, the presented solution in this letter is unique
\cite{Osborne-00}.

To analyze the topological properties of the network, 
we focus  on the adjacent matrix $A$, 
obtained from the coupling matrix
$w$ as
\begin{equation}
	A_{ij}=\begin{cases} 0 & \text{if $\hat{w}_{ji}=0$} \\ 
			1 & \text{otherwise.} \end{cases}
\label{adjacentmatrix}
\end{equation}
Hence we obtain an unweighted digraph.

\begin{figure}
\includegraphics[width=\linewidth]{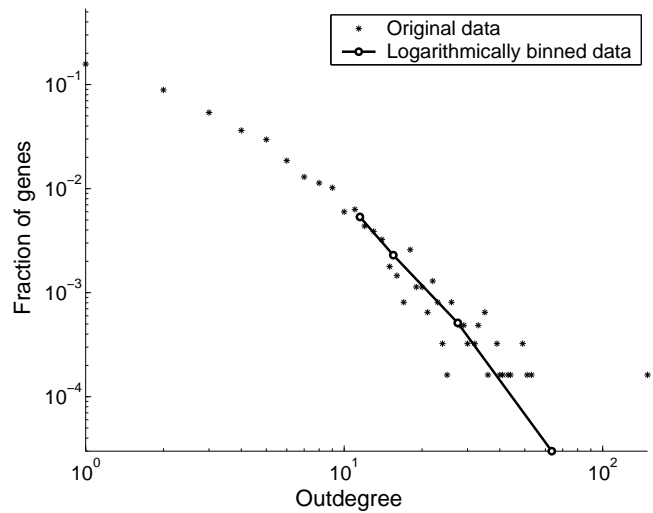}
\caption{Distribution of outdegrees for the inferred network.
	This distribution is incompatible with one from
	a random graph.
	Further, there are 3331 nodes with outdegree zero.
	\label{outdegrees}}
\end{figure}

For  this digraph, we obtain a distribution of indegrees  
varying between unity and eight.
This we attribute as an artifact of the Lasso procedure,
because the sum of the modulus of the coefficients is forced not to exceed a
specific value, and hence it is natural that there is no large spread in their
number of non-zero values.
More interesting is the distribution of outdegrees, which is depicted
in Fig.~\ref{outdegrees}.
The distribution is very skew, totally incompatible with a
Poisson distribution of a random graph \cite{Bollobas}.
For the present distribution of indegrees,
the probability that any gene gets an outdegree that exceeds 50
by accident is less than $10^{-40}$
given that the edges are independently uniformly distributed.

Our obtained distribution of outdegrees does not follow a power-law
for more than one decade,
i.e., it is not scale-free; a property that many other 
biological networks seem to have \cite{Newman-03}.
Still, the distribution is broad, and we expect it to be robust
as scale-free networks have been proven to be---both 
topologically \cite{Albert-00} and dynamically \cite{Aldana-03}.

\begin{figure}
\includegraphics[width=\linewidth]{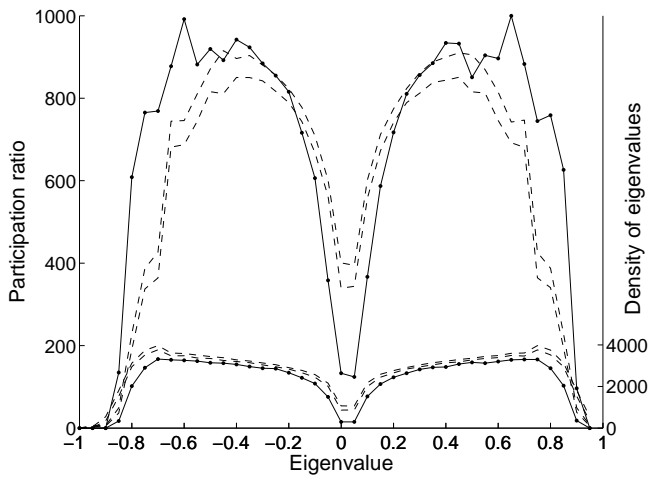}
\caption{Participation ratios  for the 
	normalized eigenvectors (upper curves) 
	and  density of eigenvalues (lower curves)
	versus eigenvalues of the transpose of the matrix $T$
	from (\ref{diffusionmatrix}).
	The results of the rewired networks, dashed curves, are 
	the variation of one standard deviation around the
	mean of 20 randomized networks
	where the degree distribution has been kept constant
	and thus comprise a null hypothesis.
	The  eigenvalues equal 	to unity, as well
	as the zero eigenvalues, have been discarded here.
	The results are binned for clarity into bins of size $0.05$.
	There is a more modular structure in the yeast network than
	could be expected from a random system.
	\label{modules}}
\end{figure}

We also search for possible modules in the obtained network.
To perform this search, we employ the formalism in 
\cite{Eriksen-03a}, where the authors unraveled modules in the internet.
Here we explore the network in undirected form, i.e., we 
study the corresponding undirected graph obtained by making
the adjacent matrix (\ref{adjacentmatrix}) symmetric.
The participation ratio of each (normalized) eigenvector 
to the transpose of the matrix 
\begin{equation}
T_{ij}=\begin{cases} 0 & \text{if nodes $i$ and $j$ not are adjacent} \\ 
			1/k_j & \text{otherwise,} \end{cases}
\label{diffusionmatrix}
\end{equation}
where $k_j$ is the degree of node $j$,
gives an estimate of the size of the corresponding module.
The eigenvalue itself corresponds to how tightly connected
the module is.
In Fig.~\ref{modules} we see how the participation ratios vary
with the eigenvalues and also the density of eigenvalues.
As a null hypothesis, we depict the variation within one standard
deviation for  values obtained from randomized networks
with the same degree distribution as the actual yeast network,
as described in \cite{BornholdtMaslov}.
We see in the figure how there are some modules
in the yeast network, 
by the fact that there are quite high participation numbers,
compared with the randomized networks,
for eigenvalues around 0.8. 
A closer analysis of these modules will be published elsewhere.

To study the biological relevance of the inferred network, we
 return to the distribution of outdegrees 
in Fig.~\ref{outdegrees}.
The gene RRN5 has the highest outdegree, 149.
According to the Saccharomyces Genome Database, SGD 
\cite{SGD}, it is involved in
transcription of rDNA by RNA poly\-me\-ra\-se I.
A systematic deletion gives an inviable organism.
The genes YHL026C and YJR079W have the second and third highest
outdegrees, 53 and 51, respectively.
According to SGD, the organism is still viable after a systematic deletion
of each.
The molecular functions of the genes are unknown, 
as are the biological processes in which they are involved.

It does not seem  too unrealistic to associate the nodes with high
outdegrees with transcription factors, 
although the edges in the obtained
network must not be interpreted as biochemical interactions only.
However, a previous study based on 273 single gene-deletion experiments for
the same kind of yeast as we have did not show any such 
correlation \cite{Featherstone-02}.
Still, though, the conjecture makes sense, and we investigate if the (known)
transcription factors of yeast are overrepresented among the genes
with highest outdegrees. 
In order to do so, we exploit the procedure proposed in \cite{Eriksen-03b}.

We rank the genes according to their outdegree, giving the highest rank
(i.e, rank number one)
to the gene with the highest outdegree.
From the GO-database \cite{GO} we obtain a classification of each gene whether
it codes for a transcription factor or not (or if it is unknown).
From these data, we form the \emph{cumulative excess} of genes 
which code for transcription factors, 
\begin{equation}
\Delta_r = \# \{ \text{TF} \le r\} \; -\; n_r  \frac{\#\{\text{TF}\}}{M},
\label{eq1}
\end{equation}
as a function of rank $r$.
All genes are ranked, so $r=1,\cdots,N$.
Here $\#\{ \text{TF} \le r\}$ is the number of genes known to code
for transcription factors and 
$n_r$ is the number of classified genes, both in the set of
genes with rank $\le r$,
$\#\{\text{TF}\}=308$ is the total number of genes known to code for 
transcription factors, and finally $M=3294$ is the total number of 
classified genes.

\begin{figure}[hbt]
\includegraphics[width=0.9\linewidth]{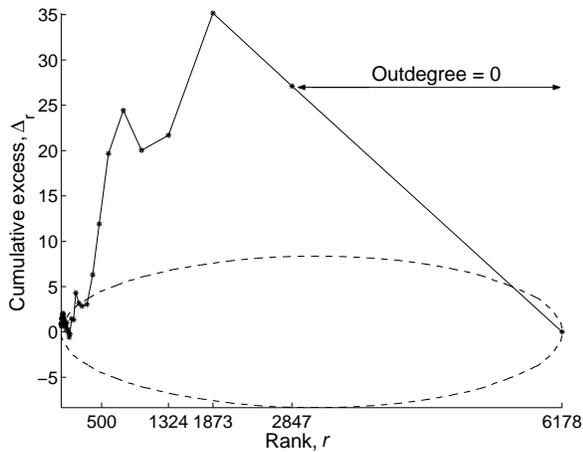}
\caption{Cumulated excess of genes coding for transcription factors,
	ranked according to their outdegrees. The dashed curves correspond
	to one standard deviation under the null hypothesis that the
	transcription factors are uniformly distributed among the
	classified genes.
	The straight lines correspond to sets of genes with the same
	outdegree, and whose order within the set thereby is arbitrary.
	\label{TF}}
\end{figure}

The number we subtract is the expected number
of genes coding for transcription factors under the null hypothesis that they
are uniformly distributed in outdegree ranks. 
In Fig.~\ref{TF} we show $\Delta_r$, 
the cumulative excess, as function of rank.
The \emph{slope} of the curve corresponds to the excess of 
transcription factors, i.e., the deviation from the null hypothesis.
\footnote{In principle, the slope is a more interesting entity than the
	cumulated excess, but it turns
	out to be less suitable for visualization \cite{Eriksen-03b}.}

The curve in Fig.~\ref{TF}
shows a clear excess of genes
coding for transcription factors among the nodes with high outdegrees,
between 400 and 2000, approximately.
To see this, we depict in  
the figure also the curves corresponding
to plus and minus one standard deviation under the null hypothesis
(dashed curves).
The ratio between the observed deviation 
and the standard deviation translate into standard Z-scores.
We have a signal of  4.8  standard deviations for the
first 737 genes,  3.2 standard deviations for the
first 1000 genes and  4.6 standard deviations for the
first 2000 genes.
Hence, we claim that the obtained distribution of genes is very
far from accidental.

In summa, we have presented the application of a specific 
inference procedure to the reverse engineering problem
of an  effective regulatory network from temporal data.
By studying the simplified network where we have
discarded the weights of the links, we find a distribution of outdegrees
which is broad, as well as modules in the network.
The existence of nodes with high outdegrees by chance is improbable.
A closer look shows that the most connected node in the network
represents a gene
which is involved in transcription.
Especially, we also find a
clear excess of genes coding for transcription factors among the
genes with high outdegrees.
Given the simplicity of our approach, utilizing only linear models and
transcript data, the method works  reasonably well.

\begin{acknowledgments}
We thank Kasper Eriksen for helpful discussions and some preliminary
datasets to analyze.
The center for industrial IT at Link\"oping university, CENIIT, 
and the Swedish research council, VR,
are acknowledged for financial support.
\end{acknowledgments}

\bibliography{referenslista}

\end{document}